\documentclass[twocolumn,showpacs,floats,floatfix,superscriptaddress,aps,pra]{revtex4}
\usepackage{amsfonts}
\usepackage{amssymb}
\usepackage{amsmath}
\usepackage{calc}
\usepackage{graphicx}
\usepackage{bm}

\def\be{ \begin{equation} }
\def\ee{ \end{equation} }
\def\bea{ \begin{eqnarray} }
\def\eea{ \end{eqnarray} }
\def\ba{ \begin{array} }
\def\ea{ \end{array} }
\def\bse{\begin{subequations}}
\def\ese{\end{subequations}}
\def\half{\tfrac12}

\def\R{\mathbf{R}}
\def\H{\mathbf{H}}
\def\O{\mathcal{O}}
\def\P{\mathbf{P}}
\def\U{\mathbf{U}}
\def\M{\mathbf{M}}
\def\Q{\mathbf{Q}}
\def\subscr{\varkappa}
\def\astate{\chi}
\def\phase{\varphi}

\def\ee{ \end{equation} }

\def\half{\tfrac12}

\def\dt{\frac{\partial}{\partial t}}

\begin{document}

\author{S. S. Ivanov}
\affiliation{Department of Physics, Sofia University, James Bourchier 5 blvd, 1164 Sofia, Bulgaria}
\author{N. V. Vitanov}
\affiliation{Department of Physics, Sofia University, James Bourchier 5 blvd, 1164 Sofia, Bulgaria}
\affiliation{Institute of Solid State Physics, Bulgarian Academy of Sciences, Tsarigradsko chauss\'{e}e 72, 1784 Sofia, Bulgaria}
\title{Steering quantum transitions between three crossing energy levels}
\date{\today }

\begin{abstract}
We calculate the propagator and the transition probabilities for a coherently driven three-state quantum system.
The energies of the three states change linearly in time, whereas the interactions between them are pulse-shaped.
We derive a highly accurate analytic approximation by assuming independent pairwise Landau-Zener transitions occurring instantly at the relevant avoided crossings, and adiabatic evolution elsewhere.
Quantum interferences are identified, which occur due to different possible evolution paths in Hilbert space between an initial and a final state.
A detailed comparison with numerical results for Gaussian-shaped pulses demonstrates a remarkable accuracy of the analytic approximation.
We use the analytic results to derive estimates for the half-width of the excitation profile, and for the parameters required for creation of a maximally coherent superposition of the three states.
These results are of potential interest in ladder climbing in alkali atoms by chirped laser pulses, in quantum rotors, in transitions between Zeeman sublevels of a $J=1$ level in a magnetic field, and in control of entanglement of a pair of spin-1/2 particles.
The results for the three-state system can be generalized, without essential difficulties, to higher dimensions.
\end{abstract}

\maketitle

%%%%%%%%%%%%%%%%%%%%%%%%%%%%%%%%%%%%%%%%%%%%%%%%%%%%%%%%%%%%%%%%%%%%%%%
%%%%%%%%%%%%%%%%%%%%%%%%%%%%%%%%%%%%%%%%%%%%%%%%%%%%%%%%%%%%%%%%%%%%%%%
\section{Introduction}
%%%%%%%%%%%%%%%%%%%%%%%%%%%%%%%%%%%%%%%%%%%%%%%%%%%%%%%%%%%%%%%%%%%%%%%
%%%%%%%%%%%%%%%%%%%%%%%%%%%%%%%%%%%%%%%%%%%%%%%%%%%%%%%%%%%%%%%%%%%%%%%
Whenever the energies of two discrete quantum states cross when plotted against some parameter, e.g. time, the transition probability is traditionally estimated by the famous Landau-Zener (LZ) formula \cite{LZ}.
Although the LZ model involves the simplest nontrivial time dependence -- linearly changing energies and a constant interaction of infinite duration,
 when applied to real physical systems with more sophisticated time dependences the LZ model often provides more accurate results than expected.
This feature (which has not been fully understood yet), and the extreme simplicity of the LZ transition probability,
 have determined the vast popularity of the LZ model, despite the availability of more sophisticated exactly soluble level-crossing models,
 e.g. the Demkov-Kunike model \cite{DK} and its special case, the Allen-Eberly-Hioe model \cite{AE}.

Numerous extensions of the LZ model to multiple levels have been proposed.
The exactly soluble multistate LZ models belong to two main types: single-crossing bow-tie models and multiple-crossings grid models.
In the \textit{bow-tie models}, where all energies cross at the same instant of time, analytic solutions have been found for three \cite{Carroll} and $N$ states \cite{Ostrovsky97,Harmin,Brundobler},
 and when one of the levels is split into two parallel levels \cite{Demkov01}.
In the \textit{grid models}, a set of $N_a$ parallel equidistant linear energies cross another set of $N_b$ such energies (Demkov-Ostrovsky model) \cite{Demkov95,Demkov95b,Usuki,Ostrovsky98}.
For $N_b=1$ (or $N_a=1$) the Demkov-Ostrovsky model reduces to the Demkov-Osherov model \cite{Demkov68,Kayanuma}.
The cases of one \cite{Kyoseva} or two \cite{Vasilev} degenerate levels have also been solved.
In the most general case of linear energies of arbitrary slopes, the general solution is not known, but exact results for some survival probabilities have been derived \cite{Shytov,Sinitsyn04,Volkov04,Volkov05}.

%For instance, the counterintuitive transitions, for which the level crossings appear in a ``wrong'' order in time, are forbidden at infinite times.
%It has been shown, though, that the probability for counterintuitive transitions is nonzero for finite interaction duration
%\cite{Rangelov} or for piecewise-linear sloped potential \cite{Yurovsky99}.

A variety of physical systems provide examples of multiple level crossings.
Among them we mention ladder climbing of atomic and molecular states by chirped laser pulses \cite{Melinger,ARPC},
 harpoon model for reactive scattering \cite{Child}, and optical shielding in cold atomic collisions \cite{shielding}.
Examples of bow-tie linkages occur, for instance, in a rf-pulse controlled Bose-Einstein condensate output coupler \cite{Mewes,magnetic}
 and in the coupling pattern of Rydberg sublevels in a magnetic field \cite{Harmin}.
A degenerate LZ model emerges when the transition between two atomic levels of angular momenta $J_{a}$ and $J_{b}=J_{a}$ or $J_{a}\pm 1$ is driven by linearly chirped laser fields of arbitrary polarizations \cite{Kyoseva,Vasilev}.

A general feature of all soluble nondegenerate multilevel crossing models is that each transition probability $P_{m\rightarrow n}$ between states
 $\psi_m$ and $\psi_n$ is given by a very simple expression, as in the original LZ model, although the derivations are not trivial.
In the grid models, in particular, the exact probabilities $P_{m\rightarrow n}$ have the same form (products of LZ probabilities
 for transition or no-transition applied at the relevant crossings) as what would be obtained by naive multiplication of LZ probabilities
 while moving across the grid of crossings from $\psi_m$ to $\psi_n$, without accounting for phases and interferences.
Quite surprisingly, interferences between different paths to the same final state, a multitude of which exist in the grid models, are not visible in the final probabilities.

In this paper we develop an analytic description of a three-state model wherein the three energies change linearly in time, with distinct slopes, thus creating three separate level crossings.
This system is particularly convenient for it presents the opportunity to investigate quantum interference through different evolution paths to the same final state,
 and in the same time, it is sufficiently simple to allow for an (approximate) analytic treatment; for the latter we use sequential two-state LZ and adiabatic-following propagators.
This system is also of practical significance for it occurs in various physical situations, for instance,
 in transitions between magnetic sublevels of a $J=1$ level \cite{magnetic},
 in chirped-pulse ladder climbing of alkali atoms \cite{rubidium},
 in rotational ladder climbing in molecules \cite{centrifuge},
 and in entanglement of a pair of spin-1/2 particles \cite{spins}.
The results provide analytic estimates of all nine transition probabilities in this system.
We do establish quantum interferences and estimate the amplitude and the frequency of the ensuing oscillation fringes, as well as the conditions for their appearance.
The analytic results also allow us to prescribe explicit recipes for quantum state engineering, for example, to create an equal, maximally coherent superposition of the three states.

This paper is organized as follows.
In Sec. \ref{Definition of the Problem} we provide the basic equations and definitions and define the problem.
In Sec. \ref{Evolution matrix} we derive the propagator, the transition probabilities and the validity conditions.
In Sec. \ref{Numerical computation vs analytical approximation} we compare our analytical approximation to numerical simulations.
Then in Sec. \ref{Applications of analytics} we demonstrate various applications of the analytics.
In Sec. \ref{Comparison with the exactly soluble Carroll-Hioe model for} we compare our model with the exactly soluble Carroll-Hioe bowtie model in the limit of vanishing static detuning.
Finally, we discuss the conclusions in Sec. \ref{Sec-conclusions}.

%%%%%%%%%%%%%%%%%%%%%%%%%%%%%%%%%%%%%%%%%%%%%%%%%%%%%%%%%%%%%%%%%%%%%%%
%%%%%%%%%%%%%%%%%%%%%%%%%%%%%%%%%%%%%%%%%%%%%%%%%%%%%%%%%%%%%%%%%%%%%%%
\section{Definition of the Problem}\label{Definition of the Problem}
%%%%%%%%%%%%%%%%%%%%%%%%%%%%%%%%%%%%%%%%%%%%%%%%%%%%%%%%%%%%%%%%%%%%%%%
%%%%%%%%%%%%%%%%%%%%%%%%%%%%%%%%%%%%%%%%%%%%%%%%%%%%%%%%%%%%%%%%%%%%%%%

%%%%%%%%%%%%%%%%%%%%%%%%%%%%%%%%%%%%%%%%%%%%%%%%%%%%%%%%%%%%%%%%%%%%%%%
\subsection{Description of the system}\label{Sec-system}

We consider a three-state system driven coherently by a pulsed external field, with the rotating-wave approximation (RWA) Hamiltonian (in units $\hbar=1$)
\begin{equation}
\label{1}
    \H(t)=\left[
    \begin{array}{ccc}
    \vspace{2mm}
    \Delta_0 + At  & \half \Omega_{12}(t)  & 0  \\
    \vspace{2mm}
    \half \Omega_{12}(t)  & 0  & \half \Omega_{23}(t)  \\
    0  & \half \Omega_{23}(t)  & \Delta_0 - At  \\
    \end{array}
\right].
\end{equation}
The diagonal elements are the (diabatic) energies (in units $\hbar$) of the three states, the second of which is taken as the zero reference point without loss of generality.
$\Delta_0$ is a static detuning, and $\pm At$ are the linearly changing terms.
To be specific, we shall use the language of laser-atom interactions, where the difference between each pair of diagonal elements is the detuning for the respective transition:
 the offset of the laser carrier frequency from the Bohr transition frequency.
The pulse-shaped functions $\Omega_{12}(t)$ and $\Omega_{23}(t)$ are the Rabi frequencies, which quantify the field-induced interactions between each pair of adjacent states,
 $\psi_1\leftrightarrow\psi_2$ and $\psi_2\leftrightarrow\psi_3$, respectively.
Each of the Rabi frequencies is proportional to the respective transition dipole moment and the laser electric-field envelope.
As evident from the zeroes in the corners of the Hamiltonian \eqref{1} we assume that the direct transition $\psi_1\leftrightarrow\psi_3$ is forbidden,
 as it occurs in free atoms when $\psi_1\leftrightarrow\psi_2$ and $\psi_2\leftrightarrow\psi_3$ are electric-dipole transitions.

The probability amplitudes of our system $\mathbf{C}(t)=\left[C_1(t),C_2(t),C_3(t)\right]^T$ satisfy the Schr\"{o}dinger equation
%%%%%%%%%%%%%%%%%
\begin{equation}
    \label{2}
    i \mathbf{\dot{C}}(t)=\H(t) \mathbf{C}(t),
\end{equation}
where the overdot denotes a time derivative.

Without loss of generality, the couplings $\Omega_{12}(t)$ and $\Omega_{23}(t)$ are assumed real and positive and, for the sake of simplicity, with the same time dependence.
For the time being the detuning $\Delta_0$ and the slope $A$ are assumed to be also positive,
%%%%%%%%%%%%%%%%%%
\begin{equation}
    \Delta_0>0~,~~A>0;
\end{equation}
we shall consider the cases of negative $\Delta_0$ and $A$ later on.
With the assumptions above, the crossing between the diabatic energies of states $\psi_1$ and $\psi_2$ occurs at time $t_-=-\tau$, where $\tau=\Delta_0/A$, between $\psi_2$ and $\psi_3$ at time $t_+=\tau$,
 and the one between $\psi_1$ and $\psi_3$ at time $t_0=0$.

\begin{figure}[tb]
\centering  \includegraphics[angle=0,width=80mm]{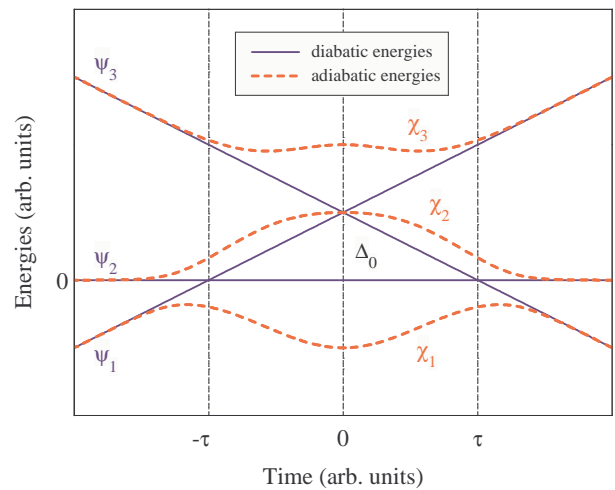}
\caption{(Color online) Diabatic and adiabatic energies vs time for a Gaussian-shaped laser pulse.
The labels denote the respective diabatic and adiabatic states.}
\label{fig1}
\end{figure}

Fig. \ref{fig1} plots diabatic and adiabatic energies vs time for a Gaussian-shaped laser pulse. We use $\psi_k$ and $\astate_k$ to denote diabatic and adiabatic states, respectively.

%==================================================================================
%==================================================================================

The objective of this paper is to find analytical expressions for the evolution matrix and for the transition probabilities between different diabatic states.

%%%%%%%%%%%%%%%%%%%%%%%%%%%%%%%%%%%%%%%%%%%%%%%%%%%%%%%%%%%%%%%%%%%%%%%
%%%%%%%%%%%%%%%%%%%%%%%%%%%%%%%%%%%%%%%%%%%%%%%%%%%%%%%%%%%%%%%%%%%%%%%
\subsection{Implementation}\label{Implementation}

The Hamiltonian \eqref{1} appears naturally in a number of specific problems of interest in time-dependent quantum dynamics of simple systems.

The first example is ladder climbing of electronic energy states in some alkali atoms, for instance, in rubidium \cite{rubidium}.
A linearly chirped laser pulse couples simultaneously both transitions 5s-5p and 5p-6s.
If the carrier frequency of the pulse is tuned on two-photon resonance with the 5s-6s transition, then the intermediate state 5p remains off resonance, by a detuning $\Delta$,
 which leads to the \textquotedblleft triangle\textquotedblright linkage pattern in Fig. \ref{fig1}.
The couplings $\Omega_{12}(t)$ and $\Omega_{23}(t)$ are the Rabi frequencies of the two transitions, which may be different (because of the different transition dipole moments)
 but have the same time dependence since they are induced by the same laser pulse.

A second example is found in rf transitions between the three magnetic sublevels $m=-1,0,1$ of a level with an angular momentum $J=1$ in an atom trapped in a magnetooptical trap.
The rf pulse provides the pulsed coupling between the $m=-1$ and $m=0$ sublevels, and also between the $m=0$ and $m=1$ sublevels.
The trapping magnetic field causes Zeeman shifts in the magnetic sublevels $m=-1$ and 1 in different directions but it does not affect the $m=0$ level \cite{magnetic}.
This linkage pattern is an example of a bowtie level crossing \cite{Carroll,Ostrovsky97,Harmin,Brundobler}.
If a quadratic Zeeman shift is taken into account, then the sublevels $m=-1$ and 1 will be shifted in the same direction,
 which will break the symmetry of the bowtie linkage and will create the \textquotedblleft triangle\textquotedblright pattern of Fig. \ref{1}.

A third example is found in quantum rotors, for instance, in rotational ladder climbing in molecules by using a pair of chirped ultrashort laser pulses \cite{centrifuge}.
The energy slope is due to the laser chirp, and the static detuning $\Delta_0$ arises due to the rotational energy splitting.
If the laser pulse duration is chosen appropriately then only three rotational states will be coupled, with their energies forming the ``triangle'' pattern of Fig. \ref{1}.

The fourth example is the entanglement between two spin-1/2 particles interacting with two crossed magnetic fields, a linear field along one axis and a pulsed field along another axis \cite{spins}.
The role of the static detuning $\Delta_0$ is played by the spin-spin coupling constant.
Three of the four collective states form a chain, which has exactly the \textquotedblleft triangle\textquotedblright\ linkage pattern of Fig. \ref{1}.
In this system, states $\psi_1$ and $\psi_3$ correspond to the product states $\left\vert \downarrow\right\rangle\left\vert \downarrow\right\rangle $
 and $\left\vert \uparrow\right\rangle\left\vert \uparrow\right\rangle $, whereas state $\psi_2$ is the entangled state
 $\left(  \left\vert \downarrow\right\rangle \left\vert \uparrow\right\rangle +\left\vert \uparrow\right\rangle \left\vert \downarrow\right\rangle \right)/\sqrt{2}$.

%%%%%%%%%%%%%%%%%%%%%%%%%%%%%%%%%%%%%%%%%%%%%%%%%%%%%%%%%%%%%%%%%%%%%%%
%%%%%%%%%%%%%%%%%%%%%%%%%%%%%%%%%%%%%%%%%%%%%%%%%%%%%%%%%%%%%%%%%%%%%%%
\section{Evolution matrix}\label{Evolution matrix}
%%%%%%%%%%%%%%%%%%%%%%%%%%%%%%%%%%%%%%%%%%%%%%%%%%%%%%%%%%%%%%%%%%%%%%%
%%%%%%%%%%%%%%%%%%%%%%%%%%%%%%%%%%%%%%%%%%%%%%%%%%%%%%%%%%%%%%%%%%%%%%%

An exact solution of the Schr\"{o}dinger equation (\ref{2}) for the Hamiltonian \eqref{1} is not known.
%As the nonadiabatic coupling is strong mainly around the crossings, we make the following assumption:
% the evolution is represented as a sequence of instantaneous two-state Landau-Zener (LZ) transitions at each crossing and adiabatic evolution elsewhere.
%Therefore we switch in adiabatic picture.
We shall derive an approximation, which is most conveniently obtained in the adiabatic basis.

%%%%%%%%%%%%%%%%%%%%%%%%%%%%%%%%%%%%%%%%%%%%%%%%%%%%%%%%%%%%%%%%%%%%%%%
\subsection{Adiabatic picture}
%%%%%%%%%%%%%%%%%%%%%%%%%%%%%%%%%%%%%%%%%%%%%%%%%%%%%%%%%%%%%%%%%%%%%%%

The adiabatic states are defined as the eigenvectors $\astate_k(t)$ ($k=1,2,3$) of the instantaneous Hamiltonian $\H(t)$.
The corresponding adiabatic amplitudes $\mathbf{A}(t)=\left[A_1(t),A_2(t),A_3(t)\right]^T$ and the diabatic ones $\mathbf{C}(t)$ are related as
%%%%%%%%%%%%%%
\begin{equation}
    \label{tr}
    \mathbf{C}(t) = \R(t)\mathbf{A}(t),
\end{equation}
%%%%%%%%%%%%%%
where $\R(t)$ is an orthogonal (because $\H(t)$ is real) transformation matrix, $\R^{-1}(t)=\R^T(t)$, whose columns are the adiabatic states $\astate_k$
 ($k=1,2,3$), with $\astate_1$ having the lowest energy and $\astate_3$ the highest energy.
As we are only interested in the populations at infinite times, we need only $\R(\pm\infty)$, rather than the explicit function $\R(t)$.
$\R(\pm\infty)$ can be easily obtained using the asymptotic behavior of $\H(t)$ at infinite times,
%%%%%%%%%%%%%%
\begin{equation}  \label{asymptotics}
    \R(-\infty)=\left[ \begin{array}{ccc}
    1  & 0  & 0  \\
    0  & 1  & 0  \\
    0  & 0  & 1  \\
    \end{array} \right],\quad
    \R(+\infty)=\left[ \begin{array}{ccc}
    0  & 0  & 1  \\
    0  & 1  & 0  \\
    1  & 0  & 0  \\
    \end{array} \right].
\end{equation}
%%%%%%%%%%%%%%

The Schr\"{o}dinger equation in adiabatic basis reads
%%%%%%%%%%%%%%%%%
\begin{equation}
    \label{4}
    i \mathbf{\dot{A}}(t)=\H_A(t) \mathbf{A}(t),
\end{equation}
%%%%%%%%%%%%%%%%
with $\H_A(t)=\R^T(t)\H(t)\R(t)-i \R^T(t) \dot{\R}(t)$, or
%%%%%%%%%%%%%%%%%
\begin{equation}
\label{5}
 \H_A(t)=\left[
    \begin{array}{ccc}
    \lambda_1(t)  & -i \nu_{12}(t)  & -i \nu_{13}(t)  \\
    -i \nu_{21}(t)  & \lambda_2(t)  & -i \nu_{23}(t)  \\
    -i \nu_{31}(t)  & -i \nu_{32}(t)  & \lambda_3(t)  \\
    \end{array}%
    \right],
\end{equation}
%%%%%%%%%%%%%%%%%
where the nonadiabatic coupling between the adiabatic states $\astate_k(t)$ and $\astate_l(t)$ is
%%%%%%%%%%%%%%%%%
\begin{equation}
    \label{6}
    \nu_{kl}(t)=\left< \astate_k(t) | \dot{\astate}_l(t) \right>=-\nu_{lk}(t).
\end{equation}
%%%%%%%%%%%%%%%%%

%In our approach we do not calculate the nonadiabatic couplings $\nu_{21}$, as we assume instantaneous two-state LZ transitions at the crossings.
%Instead, we simply obtain the nonadiabatic couplings for the

%%%%%%%%%%%%%%%%%%%%%%%%%%%%%%%%%%%%%%%%%%%%%%%%%%%%%%%%%%%%%%%%%%%%%%%
\subsection{Assumptions}
%%%%%%%%%%%%%%%%%%%%%%%%%%%%%%%%%%%%%%%%%%%%%%%%%%%%%%%%%%%%%%%%%%%%%%%

Our approach is based on two simplifying assumptions.
First, we assume that appreciable transitions take place only between neighboring adiabatic states, $\astate_1(t)\leftrightarrow \astate_2(t)$ and $\astate_2(t)\leftrightarrow \astate_3(t)$, but not between states $\astate_1(t)$ and $\astate_3(t)$, because the energies of the latter pair are split by the largest gap (cf. Fig. \ref{fig1}).
Second, we assume that the nonadiabatic transitions occur instantly at the corresponding avoided crossings and the evolution is adiabatic elsewhere.
This allows us to obtain the evolution matrix in the adiabatic basis by multiplying seven evolution matrices describing either LZ nonadiabatic transitions or adiabatic evolution.

%%%%%%%%%%%%%%%%%%%%%%%%%%%%%%%%%%%%%%%%%%%%%%%%%%%%%%%%%%%%%%%%%%%%%%%
\subsection{Evolution matrix in the adiabatic basis}
%%%%%%%%%%%%%%%%%%%%%%%%%%%%%%%%%%%%%%%%%%%%%%%%%%%%%%%%%%%%%%%%%%%%%%%

The adiabatic evolution matrix $\U^A(\infty,-\infty)$ is most conveniently determined in the adiabatic interaction representation, where the diagonal elements of $\H_A(t)$ are nullified.
The transformation to this basis reads
%%%%%%%%%%%%%%%%%
\begin{equation}
    \label{7}
    \mathbf{A}(t)=\M(t,t_0) \mathbf{B}(t),
\end{equation}
%%%%%%%%%%%%%%%%%
where
%%%%%%%%%%%%%%%%%
\bse
\bea
\label{8}
 &\M(t,t_0)=\left[
    \begin{array}{ccc}
    e^{-i\Lambda_1(t,t_0)}  & 0  & 0  \\
    0  & e^{-i\Lambda_2(t,t_0)}  & 0  \\
    0  & 0  & e^{-i\Lambda_3(t,t_0)}  \\
    \end{array}%
    \right],&\\
    \label{9a}
   & \Lambda_k(t,t_0)=\int_{t_0}^t \lambda_k(t^\prime)d t^\prime,&
\eea
\ese
%%%%%%%%%%%%%%%%%
and $t_0$ is an arbitrary fixed time.
The Schr\"{o}dinger equation in this basis reads
%%%%%%%%%%%%%%%%%
\begin{equation}
    \label{10}
    i \mathbf{\dot{B}}(t)=\H_B(t) \mathbf{B}(t),
\end{equation}
%%%%%%%%%%%%%%%%
with
%%%%%%%%%%%%%%%%%
\begin{equation}
\label{11}
 \H_B(t)=-i\left[
    \begin{array}{ccc}
    0  & \nu_{12}e^{i\Lambda_{12}(t,t_0)}  & \nu_{13}e^{i\Lambda_{13}(t,t_0)}  \\
    \nu_{21}e^{i\Lambda_{21}(t,t_0)}  & 0  & \nu_{23}e^{i\Lambda_{23}(t,t_0)}  \\
    \nu_{31}e^{i\Lambda_{31}(t,t_0)}  & \nu_{32}e^{i\Lambda_{32}(t,t_0)}  & 0  \\
    \end{array}%
    \right],
\end{equation}
%%%%%%%%%%%%%%%%%
where $\Lambda_{kl}(t,t_0) \equiv \Lambda_k(t,t_0)-\Lambda_l(t,t_0)$.
In this basis the propagator for adiabatic evolution is the identity matrix.

The LZ transitions at the crossings at times $-\tau,0,\tau$ are described by the transition matrices \cite{Rangelov}
%%%%%%%%%%%%%%%%%
\begin{subequations}
\bea
\label{12a}
 &\U_{LZ}(-\tau)=\left[
    \begin{array}{ccc}
    \sqrt{q_-}e^{-i\phi_-}  & -\sqrt{p_-}            & 0  \\
    \sqrt{p_-}              & \sqrt{q_-}e^{i\phi_-}  & 0  \\
    0                       & 0                      & 1  \\
    \end{array}%
    \right],&\\
\label{12c}
& \U_{LZ}(0)=\left[
    \begin{array}{ccc}
    1              & 0                       & 0                      \\
    0              & \sqrt{q_0}e^{-i\phi_0}  & -\sqrt{p_0}            \\
    0              & \sqrt{p_0}              & \sqrt{q_0}e^{i\phi_0}  \\
    \end{array}%
    \right],&\\
\label{12b}
& \U_{LZ}(\tau)=\left[
    \begin{array}{ccc}
    \sqrt{q_+}e^{-i\phi_+}  & -\sqrt{p_+}            & 0  \\
    \sqrt{p_+}              & \sqrt{q_+}e^{i\phi_+}  & 0  \\
    0                       & 0                      & 1  \\
    \end{array}%
    \right],&
\eea
\end{subequations}
%%%%%%%%%%%%%%%%%
where $p_{\subscr}$ ($\subscr=-,0,+$) is the LZ probability of nonadiabatic transition and $q_{\subscr}$ is the no-transition probability at the crossings at times $-\tau,0,\tau$,
%%%%%%%%%%%%%%%%%
\begin{equation}
    \label{13}
    p_{\subscr}=e^{-\pi a_{\subscr}^2}, \quad q_{\subscr}=1-p_{\subscr}.
\end{equation}
%%%%%%%%%%%%%%%%%
Here
%%%%%%%%%%%%%%%%%
\begin{subequations}
\bea
    \label{14}
&&    a_-=\Omega_{12}(-\tau)/(2A)^{1/2},\\
&&    a_0=\Omega_{eff}(0)/2A^{1/2},\\
&&    a_+=\Omega_{23}(\tau)/(2A)^{1/2},\\
&&    \phi_{\subscr}=\mathrm{arg}~\Gamma (1-i a_{\subscr}^2)+\frac{\pi}{4}+a_{\subscr}^2(\ln a_{\subscr}^2-1),
\eea
\end{subequations}
%%%%%%%%%%%%%%%%%
where $\Omega_{eff}(0)$ is the effective coupling between states $\psi_1$ and $\psi_3$ at crossing time $t=0$;
 it is determined by the splitting between the adiabatic curves $\lambda_2(t)$ and $\lambda_3(t)$,
%%%%%%%%%%%%%%%%%
\begin{equation}
    \label{15}
    \Omega_{eff}(0) = \lambda_2(0)-\lambda_3(0) =\half\left(-\Delta_0+\sqrt{\Delta_0^2+2\Omega_0^2}\right).
\end{equation}
%%%%%%%%%%%%%%%%%

The propagator in the adiabatic basis reads
\bea\label{adiabatic propagator}
\U^A(\infty,-\infty)=\M(\infty,\tau)\U_{LZ}(\tau)\M(\tau,0)\U_{LZ}(0)\notag\\
\times \M(0,-\tau)\U_{LZ}(-\tau)\M(-\tau,-\infty).
\eea

%%%%%%%%%%%%%%%%%%%%%%%%%%%%%%%%%%%%%%%%%%%%%%%%%%%%%%%%%%%%%%%%%%%%%%%
\subsection{Propagator and transition probabilities in the diabatic basis}
%%%%%%%%%%%%%%%%%%%%%%%%%%%%%%%%%%%%%%%%%%%%%%%%%%%%%%%%%%%%%%%%%%%%%%%

Below we present the diabatic propagator in an explicit form.
For simplicity, we assume equal couplings
%%%%
\begin{equation}
\label{equal}
\Omega_{12}(t) = \Omega_{23}(t) = \Omega(t),
\end{equation}
%%%%
although our approach is valid in the general case.
This constraint is not applicable for the ladder climbing system, considered in Sec. \ref{Implementation}, where the couplings are naturally different due to the different transition dipole moments,
 but is still valid for the other systems discussed.
Then $\Lambda_{kl}(0,-t)=\Lambda_{kl}(t,0)$, $a_+=a_-=a$, $\phi_+=\phi_-=\phi$, $p_+=p_-=p$, and $q_+=q_-=q$.

We find the propagator in the original diabatic basis by using Eqs.~\eqref{tr}, \eqref{asymptotics} and \eqref{adiabatic propagator} as
 $\U(\infty,-\infty)=\R(\infty)\U^A(\infty,-\infty)\R^T(-\infty)$, or explicitly,
%%%%%%%%%%%%%%%%%
\begin{widetext}
\begin{equation}
\label{propagator}
 \U(\infty,-\infty)=\left[
    \begin{array}{ccc}
    e^{i\phase_1+i\phase_3}\sqrt{p p_0}
    & e^{i\phi+i\phase_2+i\phase_3}\sqrt{q p_0}
    & e^{i\phi_0+2i\phase_3}\sqrt{q_0}  \\
    e^{-i\phi-i\phase_1+i\phase_2}\sqrt{pq}+e^{i\phi-i\phi_0+i\phase_1+i\phase_2}\sqrt{pq q_0}
    & -e^{-2i\phase_1+2i\phase_2}p+e^{2i\phi-i\phase_0+2i\phase_2}q\sqrt{q_0}
    & -e^{i\phi+i\phase_2+i\phase_3}\sqrt{q p_0}  \\
    e^{-2i\phi}q-e^{-i\phi_0+2i\phase_1}p\sqrt{q_0}
    & -e^{-i\phi-i\phase_1+i\phase_2}\sqrt{pq}-e^{i\phi-i\phi_0+i\phase_1+i\phase_2}\sqrt{pqq_0}
    & e^{i\phase_1+i\phase_3}\sqrt{p p_0}  \\
    \end{array}
    \right],
\end{equation}
%%%%%%%%%%%%%%%%%
with $\phase_1=\Lambda_{12}(\tau,0)$, $\phase_2=\Lambda_{12}(\infty,0)$, $\phase_3=\Lambda_{13}(\infty,0)$.
The transition probability matrix, i.e. the matrix of the absolute squares of the elements of the propagator \eqref{propagator}, reads
%%%%%%%%%%%%%%%%%
\begin{equation}\label{probabilities}
 \P=\left[
    \begin{array}{ccc}
    p p_0
    & q p_0
    & q_0  \\
    qp+pq_0q+2qp\sqrt{q_0}\cos{\gamma}
    & p^2+q^2q_0-2qp\sqrt{q_0}\cos{\gamma}
    & q p_0  \\
    q^2+q_0p^2-2qp\sqrt{q_0}\cos{\gamma}
    & qp+pq_0q+2qp\sqrt{q_0}\cos{\gamma}
    & p p_0  \\
    \end{array}%
    \right],
\end{equation}
\end{widetext}
%%%%%%%%%%%%%%%%%
where
%%%%%%%%%%%%%%%%%
\begin{equation}
    \label{18after2}
    \gamma=2\phi-\phi_0+2\phase_1.
\end{equation}

The element at the $m$-th row and the $n$-th column of the matrix \eqref{probabilities} is the transition probability $P_{n\rightarrow m}$,
 that is the population of state $m$ at infinite time, when the system starts in state $n$ in the infinite past.
The survival probabilities $P_{1\rightarrow 1}$ and $P_{3\rightarrow 3}$ coincide with the exact expressions conjectured \cite{Brundobler} and derived exactly for constant couplings \cite{Shytov,Volkov04} earlier.

In Eq.~(\ref{probabilities}) we recognize interference terms, which arise because of the availability of two alternative propagating paths in the Hilbert space.
There is also a symmetry with respect to the skew diagonal due to the equal couplings between neighboring states \eqref{equal} and the equal (in magnitude) slopes of the energies of states $\psi_1$ and $\psi_3$ \eqref{1}.

%%%%%%%%%%%%%%%%%%%%%%%%%%%%%%%%%%%%%%%%%%%%%%%%%%%%%%%%%%%%%%%%%%%%%%%
\subsection{Conditions of validity}
%%%%%%%%%%%%%%%%%%%%%%%%%%%%%%%%%%%%%%%%%%%%%%%%%%%%%%%%%%%%%%%%%%%%%%%

As already stressed, our approach presumes that the nonadiabatic transitions occur in well-separated confined time intervals.
This means that the characteristic transition times are shorter than the times between the crossings, or $t_{transition}\lesssim \tau$.
The transition times for diabatic ($\Omega^2\ll A$) and adiabatic ($\Omega^2\gg A$) regimes are \cite{LZ times}
%%%%%%%%%%%%%%%%%
%\begin{eqnarray}
%    \label{23a}
%    t_{transition}\approx \sqrt{2\pi} \hspace{3mm} \mathrm{diabatic~ regime}   \cr
%    t_{transition}\approx 2\frac{\Omega}{A^{1/2}} \hspace{3mm} \mathrm{adiabatic~ regime}
%\end{eqnarray}
%%%%%%%%%%%%%%%%%
\bse
\bea    \label{23}
    &&t_{transition}\approx \sqrt{2\pi/A},\quad \mathrm{~diabatic~ regime},\\
    &&t_{transition}\approx 2\Omega/A,\qquad \mathrm{adiabatic~ regime}.
\eea
\ese
%%%%%%%%%%%%%%%%%
This leads to the following conditions for validity:
%%%%%%%%%%%%%%%%%
\bse    \label{conditions}
\bea
    &&\Delta_0\gtrsim \sqrt{2\pi A},\quad \mathrm{diabatic~ regime},\\
    &&\Delta_0\gtrsim 2\Omega,\quad\quad \mathrm{~adiabatic~ regime}.
\eea
\ese
%%%%%%%%%%%%%%%%%

We shall demonstrate that the LZ-based approximation \eqref{probabilities} outperforms its formal conditions of validity \eqref{conditions} and is valid beyond the respective ranges.

%%%%%%%%%%%%%%%%%%%%%%%%%%%%%%%%%%%%%%%%%%%%%%%%%%%%%%%%%%%%%%%%%%%%%%%
\subsection{Case of $\Delta_0<0$ and/or $A<0$}
%%%%%%%%%%%%%%%%%%%%%%%%%%%%%%%%%%%%%%%%%%%%%%%%%%%%%%%%%%%%%%%%%%%%%%%

Above we assumed that $\Delta_0>0$ and $A>0$.
Now we consider the cases $\Delta_0<0$ and $A<0$.
We assume that the couplings are even functions, $\Omega(-t)=\Omega(t)$.

\emph{Negative static detuning} ($\Delta_0<0$).

%We first notice that $\H(-\Delta_0,-t)=-\H(\Delta_0,t)$.
The Schr\"{o}dinger equation for the propagator $\U(\Delta_0;t,t_i)$ is
%%%%%%%%%%%%%%%%%
\be      \label{19a}
      i \dt \U(\Delta_0;t,t_i) = \H(\Delta_0,t) \U(\Delta_0;t,t_i).
\ee
By changing the signs of $\Delta_0$, $t$ and $t_i$ in Eq \eqref{19a}, we obtain the same equation, but with the $\Omega(t)$ replaced by $-\Omega(t)$ [see Eq.~\eqref{1}].
It is easy to see that the change of sign of $\Omega(t)$ is equivalent to the transformation $\U\rightarrow \U^\prime=\Q\U\Q$ where $\Q$ is the diagonal matrix $\Q=\mathrm{diag}\{1,-1,1\}$.
Hence we find
\be      \label{19b}
      i \dt \U^\prime(-\Delta_0;-t,-t_i) = \H(\Delta_0,t) \U^\prime(-\Delta_0;-t,-t_i).
\ee
%%%%%%%%%%%%%%%%
Because the initial condition at $t\rightarrow -\infty$ for $\U(\Delta_0;t,t_i)$ and $\U^\prime(-\Delta_0;-t,-t_i)$ at $t\rightarrow -\infty$ is the same,
%%%%%%%%%%%%%%%%%
  \begin{equation} \label{20a}
      \U(\Delta_0;-\infty,-\infty)=\U^\prime(-\Delta_0;\infty,\infty)=\mathbf{I},
  \end{equation}
%%%%%%%%%%%%%%%%%
we conclude that $\U(\Delta_0;t,t_i)=\U^\prime(-\Delta_0;-t,-t_i)$; hence
%%%%%%%%%%%%%%%%%
\bea \label{20b}
     \U(-\Delta_0;\infty,-\infty) &=& \Q\U(\Delta_0;-\infty,\infty)\Q \notag \\
 &=& \Q\U(\Delta_0;\infty,-\infty) ^\dagger \Q.
\eea
%%%%%%%%%%%%%%%%%
Therefore
%%%%%%%%%%%%%%%%%
  \begin{equation}
      \label{21}
      P_{m\rightarrow n}(-\Delta_0)=P_{n\rightarrow m}(\Delta_0),\quad (m,n=1,2,3).
\end{equation}
%%%%%%%%%%%%%%%%%

\emph{Negative chirp rate} ($A<0$).

We notice that $H_{11}(A)=H_{33}(-A)$, i.e. the change of sign of $A$ is equivalent to exchanging the indices 1 and 3.
Hence the probabilities for $A<0$ are obtained from these for $A>0$ using the relation
%%%%%%%%%%%%%%%%%
%\begin{subequations}
%  \begin{equation}
%      \label{22a}
%      p_{1n}(A)=p_{3n}(-A),
%  \end{equation}
%  \begin{equation}
%      \label{22b}
%      p_{n1}(A)=p_{n3}(-A).
%  \end{equation}
%\end{subequations}
  \begin{equation}   \label{22}
      P_{m\rightarrow n}(-A)=P_{4-m \rightarrow 4-n}(A),\quad \left(m,n=1,2,3\right).
  \end{equation}
%%%%%%%%%%%%%%%%

%%%%%%%%%%%%%%%%%%%%%%%%%%%%%%%%%%%%%%%%%%%%%%%%%%%%%%%%%%%%%%%%%%%%%%%
%%%%%%%%%%%%%%%%%%%%%%%%%%%%%%%%%%%%%%%%%%%%%%%%%%%%%%%%%%%%%%%%%%%%%%%
\section{Comparison of analytical and numerical results}\label{Numerical computation vs analytical approximation}
%%%%%%%%%%%%%%%%%%%%%%%%%%%%%%%%%%%%%%%%%%%%%%%%%%%%%%%%%%%%%%%%%%%%%%%
%%%%%%%%%%%%%%%%%%%%%%%%%%%%%%%%%%%%%%%%%%%%%%%%%%%%%%%%%%%%%%%%%%%%%%%

Below we compare our analytical approximation with numerical simulations.
We take for definiteness the couplings in Eq.~(\ref{1}) to be Gaussians, $\Omega(t)=\Omega_0 e^{-t^2/T^2}$.
%We present the population $P_{m\rightarrow n}$ as a function of a varied parameter.

%==================================================================================
\begin{figure}[tb]
\centering  \includegraphics[angle=0,width=80mm]{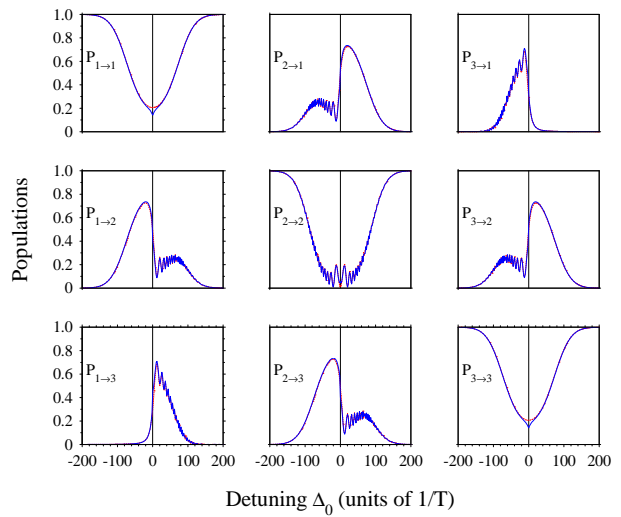}
\caption{(Color online)
The transition probabilities $P_{m\rightarrow n}$ for the transition $m\rightarrow n$ vs the detuning $\Delta_0$ for $A=100/T^2,~\Omega_0=10/T$.
Each frame compares the numerical (dashed red) and analytical (solid blue) results.}
\label{fig2}
\end{figure}
%==================================================================================

Figure \ref{fig2} shows the nine transition probabilities vs the static detuning $\Delta_0$.
An excellent agreement is observed between analytics and numerics, which are barely discernible.
This agreement indicates that the dynamics is indeed driven by separated level-crossing transitions of LZ type.
The analytic approximation \eqref{probabilities} is clearly valid beyond its formal range of validity, defined by conditions \eqref{conditions},
 which suggest $|\Delta_0| \gtrsim 25/T$ for the parameters in this figure.
The figure also demonstrates that the detuning can be used as a control parameter for the probabilities in wide ranges.
%Because of the accuracy of the analytic approximation, for any desired combination of probabilities, the values of the control parameter -- the detuning $\Delta_0$ -- can be determined analytically, including for estimating the feasibility of such a combination.

For $\Delta_0>0$ the five probabilities on the first row and the last column vary smoothly, in agreement with the analytic prediction.
The two-photon probability $P_{3\rightarrow 1}$ vanishes rapidly with $\Delta_0$, as expected, at a much faster pace than the other probabilities.
The other four probabilities $P_{1\rightarrow 2}$, $P_{1\rightarrow 3}$, $P_{2\rightarrow 2}$ and $P_{2\rightarrow 3}$ exhibit oscillations, in agreement with the analytic prediction,
 due to the existence of two alternative paths of different length from the initial to the final state (see Fig. \ref{fig1}), with an ensuing interference.
It is noteworthy that these oscillations, due to path interference, are not particularly pronounced, which might be a little surprising at first glance.
However, a more careful analysis reveals that when a control parameter is varied, such as the static detuning $\Delta_0$ here,
 it changes not only the relative phase along the two paths (which causes the oscillations), but also the LZ probabilities $p_\subscr$ and $q_\subscr$ ($\subscr=-,0,+$).
Indeed, as $\Delta_0$ increases, we have $p_\pm\rightarrow 1$ because the crossings at times $\pm \tau$ move away from the center of the pulses and $\Omega(\pm\tau)\rightarrow 0$.
These probabilities affect both the average value of $P_{m\rightarrow n}$ and the oscillation amplitude,
 with $P_{m\rightarrow n}$ tending eventually to either 0 or 1 for large $\Delta_0$, while the oscillation amplitude (which is proportional to $p_\pm$) is damped.

Similar conclusions apply to the case of $\Delta_0<0$ because of the symmetry property \eqref{21}.
It is easy to see from here that the survival probabilities $P_{n\rightarrow n}$ ($n=1,2,3$) are symmetric vs $\Delta_0$, as indeed seen in Fig. \ref{fig2}.

%==================================================================================
\begin{figure}[tb]
\centering  \includegraphics[angle=0,width=80mm]{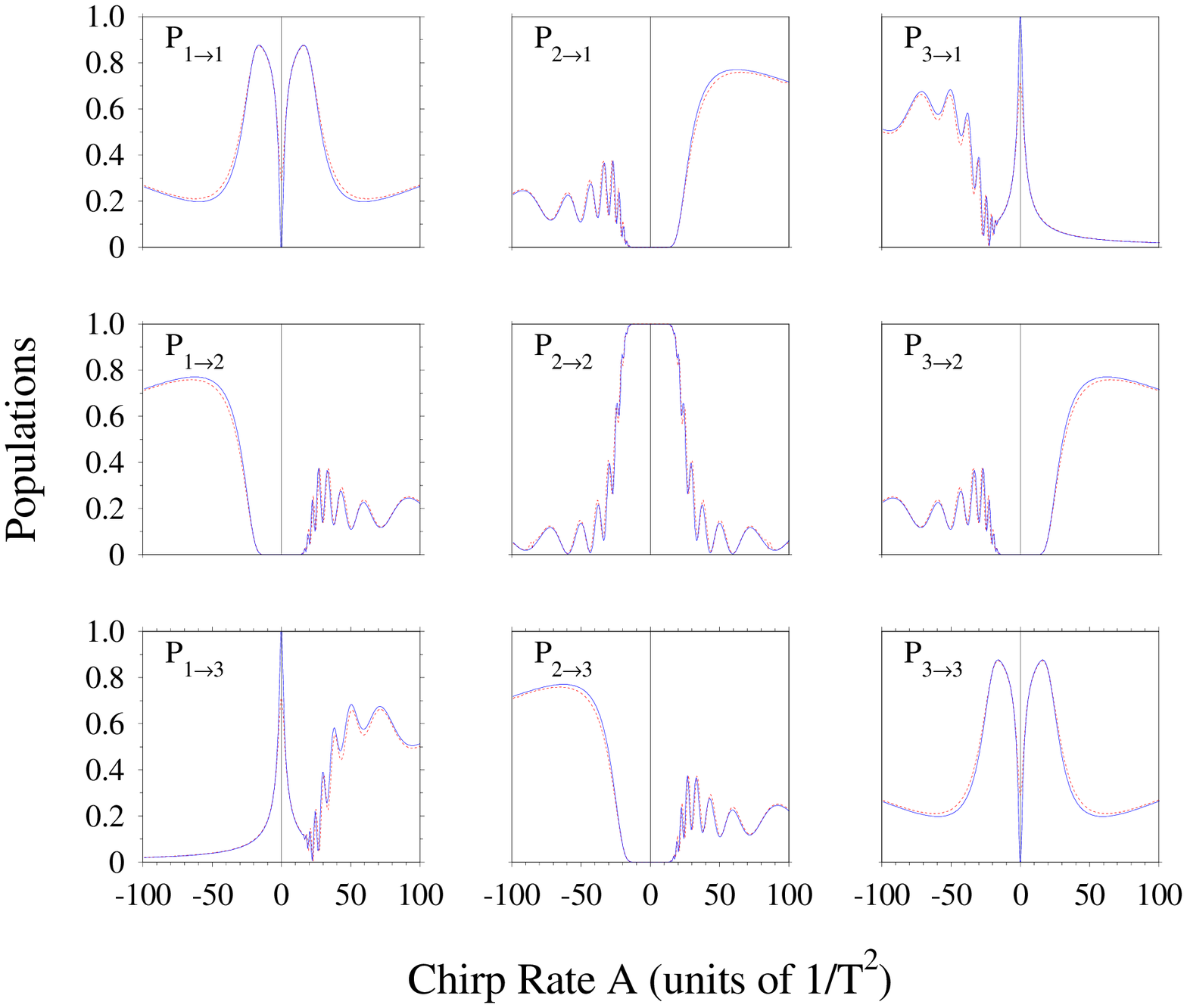}
\caption{(Color online)
The transition probabilities $P_{m\rightarrow n}$ for the transition $m\rightarrow n$ vs the energy slope $A$ for $\Delta_0=30/T,~\Omega_0=10/T$.
Each frame compares the numerical (dashed red) and analytical (solid blue) results.}
\label{fig3}
\end{figure}
%==================================================================================

Figure \ref{fig3} displays the transition probabilities vs the chirp rate $A$.
An excellent agreement is again observed between analytics and numerics.
We have verified that the analytic approximation \eqref{probabilities} is valid well beyond its formal range of validity conditions \eqref{conditions},
 which suggest $|A|\lesssim 140/T^2$ for this figure; this is not shown because our intention here is to show the small-$A$ range that exhibits interference patterns.
As with the static detuning in Fig.~\ref{fig2}, this figure demonstrates the symmetry with respect to the sign inversion of $A$, derived in Eq.~\eqref{22}:
 the change $A\rightarrow -A$ is equivalent to the exchange of the indices 1 and 3.
The observed additional symmetry, $P_{2\rightarrow 1} \equiv P_{3\rightarrow 2}$ and $P_{1\rightarrow 2} \equiv P_{2\rightarrow 3}$,
 is a consequence from the assumptions of equal Rabi frequencies and equal (in magnitude) slopes of the energies of states $\psi_1$ and $\psi_3$.
The figure also shows that, with the exception of the survival probabilities $P_{n\rightarrow n}$ $(n=1,2,3)$,
 all other probabilities are asymmetric vs the chirp rate $A$, unlike the two-state level-crossing case.
For $A>0$, as for $\Delta_0>0$ in Fig.~\ref{fig2}, oscillations are observed in the four probabilities in the lower left corner but not for the probabilities in the top row and the right column.
On the contrary, for $A<0$, oscillations are observed only in the four probabilities in the top right corner.
As discussed in regard to Fig.~\ref{fig2}, the observation of these oscillations is in full agreement with their interpretation
 as resulting from interference between two different evolution paths to the relevant final state.

Like the static detuning $\Delta_0$, the energy slope $A$ can be used as a control parameter because it affects the probabilities considerably.
Around the origin ($A=0$) the system is in adiabatic regime, while for large $|A|$ it is in diabatic regime.
For instance, when the system is initially in $\psi_1$, around the origin ($A=0$) the population flows mostly into state $\psi_3$, following the adiabatic state $\astate_1(t)$.
On the contrary, for large $A$ it eventually returns to $\psi_1$ (not visible for the chirp range in Fig. \ref{fig3}).

%==================================================================================
\begin{figure}[tb]
\centering  \includegraphics[angle=0,width=80mm]{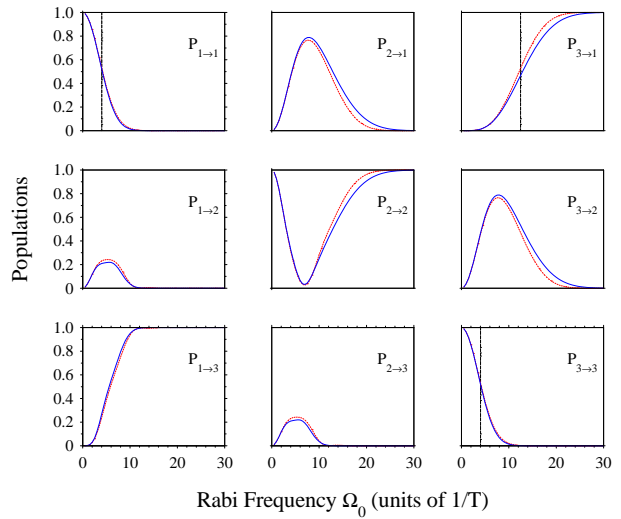}
\caption{(Color online)
The transition probabilities $P_{m\rightarrow n}$ for the transition $m\rightarrow n$ vs the Rabi frequency $\Omega_0$ for $\Delta_0=10/T,~A=30/T^2$.
Each frame compares the numerical (dashed red) and analytical (solid blue) results.
The vertical dashed lines for $P_{1\rightarrow 1}$, $P_{3\rightarrow 3}$ and $P_{3\rightarrow 1}$ show the values $\Omega_{1/2}$ of the Rabi frequency for half population in the relevant states,
 predicted by our model, Eqs.~\eqref{25a} and \eqref{25c}.}
\label{fig4}
\end{figure}
%==================================================================================

Diabatic and adiabatic regimes are easy to identify also in Fig. \ref{fig4}, where the nine probabilities are plotted vs the peak Rabi frequency $\Omega_0$, which is another control parameter.
Consider our system initially prepared in state $\psi_1$.
For weak couplings the system evolves diabatically and therefore it is most likely to end up in the same state $\psi_1$.
As the couplings increase, the system switches gradually from diabatic to adiabatic evolution;
 for strong couplings the evolution proceeds along the adiabatic state $\astate_1(t)$, and we observe nearly complete population transfer to state $\psi_3$.

Returning to the issue of oscillations, such are barely seen in Fig. \ref{fig4}.
As discussed in relation to Fig. \ref{fig2}, a varying control parameter changes, besides the relative phase of the interfering paths, also the probabilities $p_\subscr$ and $q_\subscr$,
 which eventually acquire their asymptotic values of 0 or 1; in these limits the oscillations vanish.
The probabilities depend on the peak Rabi frequency $\Omega_0$ much more sensitively than on the static detuning $\Delta_0$ and the energy slope $A$;
 consequently, clear oscillations are seen vs $\Delta_0$ and $A$, but not vs $\Omega_0$, because the dependence of $p_\subscr$ on $\Omega_0$ is strongest (essentially Gaussian),
 and hence the approach to the asymptotic values of the probabilities is fastest.

%%%%%%%%%%%%%%%%%%%%%%%%%%%%%%%%%%%%%%%%%%%%%%%%%%%%%%%%%%%%%%%%%%%%%%%
%%%%%%%%%%%%%%%%%%%%%%%%%%%%%%%%%%%%%%%%%%%%%%%%%%%%%%%%%%%%%%%%%%%%%%%
\section{Applications of analytics}\label{Applications of analytics}
%%%%%%%%%%%%%%%%%%%%%%%%%%%%%%%%%%%%%%%%%%%%%%%%%%%%%%%%%%%%%%%%%%%%%%%

In this section we shall use our analytic approximation for the transition probabilities \eqref{probabilities} to derive several useful properties of the triple-crossing system.

\subsection{Analytical linewidth}
%%%%%%%%%%%%%%%%%%%%%%%%%%%%%%%%%%%%%%%%%%%%%%%%%%%%%%%%%%%%%%%%%%%%%%%

We begin by deriving approximate expressions for the Rabi frequency required to reach $50\%$ population in the $n$-th state for the transition $m\rightarrow n$.
Simple expressions are found for the transition $3\rightarrow 1$,
%%%%%%%%%%%%%%%%%
\begin{equation}   \label{25a}
      \Omega_{1/2} =2\sqrt{\frac{2A\ln 2+\Delta_0\sqrt{\pi A \ln2}}{\pi}},
\end{equation}
%%%%%%%%%%%%%%%%%
and for the transitions $1\rightarrow 1$ and $3\rightarrow 3$,
\bea  \label{25c}
      \Omega_{1/2} &=& \frac{2}{\left(\alpha+4\right)}
      \left[ \frac{2A\alpha(\alpha+4)\ln2}{\pi} -\alpha\Delta_0^2\right. \notag\\
	&&+\alpha \Delta_0 \left.\sqrt{\Delta_0^2 + \frac{A\alpha(\alpha+4)\ln2}{\pi}}\right]^{\half},
\eea
%%%%%%%%%%%%%%%%%
where $\alpha=\exp\left(2\Delta_0^2/A^2\right)$.
These values are indicated by vertical lines in Fig.~\ref{fig4} and are seen to be in excellent agreement with the exact values.
%The analytic formulas for the values of $\Omega_{1/2}$ for the other transitions are too cumbersome to be presented here.

%%%%%%%%%%%%%%%%%%%%%%%%%%%%%%%%%%%%%%%%%%%%%%%%%%%%%%%%%%%%%%%%%%%%%%%
\subsection{Creation of superpositions}
%%%%%%%%%%%%%%%%%%%%%%%%%%%%%%%%%%%%%%%%%%%%%%%%%%%%%%%%%%%%%%%%%%%%%%%

%==================================================================================
\begin{figure}[tb]
\centering  \includegraphics[angle=0,width=65mm]{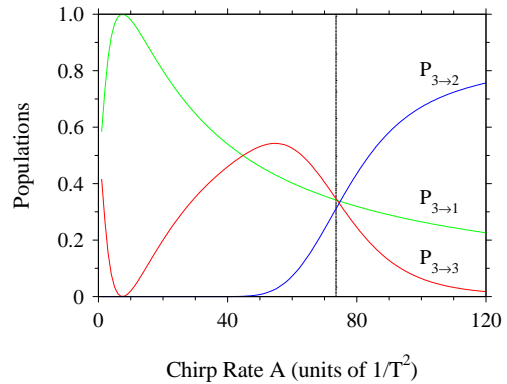}
\caption{(Color online)
The final populations of states $\psi_1$, $\psi_2$ and $\psi_3$ vs the chirp rate $A$ for fixed $\Delta_0=100/T$ and $\Omega_0=36.2/T$, provided the system is initially in state $\psi_3$.
The three curves cross at about $A\approx 74.5/T^2$, indicating the creation of a maximally coherent superposition with populations $P_1=P_2=P_3=1/3$,
 which is very close to the solution of Eqs.~\eqref{coha} and \eqref{cohb}, $A=73.6/T^2$, shown with a vertical dashed line.}
\label{fig5}
\end{figure}
%==================================================================================

If we prepare our system initially in state $\psi_1$ and use $A<0$, or in state $\psi_3$ and use $A>0$, it is possible to determine by means of our analytical model values of $\Delta_0$, $A$ and $\Omega_0$,
 so that we achieve arbitrary preselected populations at the end.
For example, for a maximally coherent superposition state, i.e. $P_1=P_2=P_3=\frac13$, we need $p=\frac{1}{2}$ and $p_0=\frac{2}{3}$.
This yields the following set of equations for $\Delta_0$, $\Omega_0$, and $A$:
%%%%%%%%%%%%%%%%%
\bse\bea
      \label{coha}
 &     \frac{1}{2} e^{2\Delta_0^2/A^2}\ln2-\Delta_0\sqrt{\frac{\pi\ln 3/2}{A}}-2\ln 3/2=0, & \\
      \label{cohb}
 &     \Omega_0=\sqrt{\frac{2A\ln2}{\pi}} e^{\Delta_0^2/A^2}.&
\eea\ese
%%%%%%%%%%%%%%%%%

An example is shown in Fig.~\ref{fig5} where the three final probabilities $P_{3\rightarrow 1}$, $P_{3\rightarrow 2}$ and $P_{3\rightarrow 3}$ are plotted versus the chirp rate $A$.
The three probabilities cross (indicating the creation of a maximally coherent superposition state) approximately at the value predicted by Eqs.~\eqref{coha} and \eqref{cohb}, shown by the vertical line.

%This result may be used to produce a maximal coherent superposition of Zeeman sublevels of a $J=1$ level in a magnetic field with very high fidelity.

%%%%%%%%%%%%%%%%%%%%%%%%%%%%%%%%%%%%%%%%%%%%%%%%%%%%%%%%%%%%%%%%%%%%%%%
%%%%%%%%%%%%%%%%%%%%%%%%%%%%%%%%%%%%%%%%%%%%%%%%%%%%%%%%%%%%%%%%%%%%%%%
\section{Comparison with the exactly soluble Carroll-Hioe model for $\Delta_0=0$}\label{Comparison with the exactly soluble Carroll-Hioe model for}
%%%%%%%%%%%%%%%%%%%%%%%%%%%%%%%%%%%%%%%%%%%%%%%%%%%%%%%%%%%%%%%%%%%%%%%
%%%%%%%%%%%%%%%%%%%%%%%%%%%%%%%%%%%%%%%%%%%%%%%%%%%%%%%%%%%%%%%%%%%%%%%

For $\Delta_0=0$ and constant couplings, the Hamiltonian \eqref{1} allows for an exact solution -- this is the Carroll-Hioe (CH) bowtie model \cite{Carroll}.
The transition probability matrix for the CH model reads
\be\label{CH}
\P_{CH} = \left[ \begin{array}{ccc}
p_{c}^{2} & 2p_{c}\left( 1-p_{c}\right)  & \left( 1-p_{c}\right) ^{2} \\
2p_{c}\left( 1-p_{c}\right)  & \left( 1-2p_{c}\right) ^{2} & 2p_{c}\left( 1-p_{c}\right)  \\
\left( 1-p_{c}\right) ^{2} & 2p_{c}\left( 1-p_{c}\right)  & p_{c}^{2}
\end{array}\right],
\ee
where
\be
p_c=e^{-\pi a^2/2},\quad a=\Omega/\sqrt{2A}.
\ee

We use this exact result as a reference for the $\Delta_0=0$ limit of our approximate method, applied for constant coupling $\Omega(t)=\Omega=const$. We emphasize that taking this limit is an abuse of the method because in the derivation we have assumed that the crossings are {\em separated}, which has justified the multiplication of propagators.
Nonetheless, it is curious and instructive to push our approximation to this limit.
For $\Delta_0=0$ the LZ parameters are $a_\pm=\Omega/\sqrt{2A}=a$ and $a_0=a/2$.
Therefore we have $p_0^4=p_c^2=p$.

%==================================================================================
\begin{figure}[tb]
\centering  \includegraphics[angle=0,width=80mm]{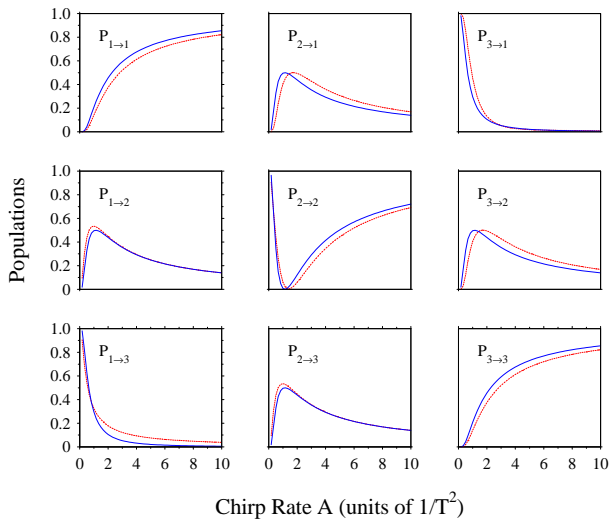}
\caption{(Color online)
Comparison of the probabilities \eqref{CH} in the exactly soluble Carroll-Hioe model (dashed red line) with our approximate solution \eqref{probabilities} (solid blue line)
 for $\Delta_0=0$ as functions of the chirp rate $A$.
Here $\Omega=1/T$.}
\label{fig6}
\end{figure}
%==================================================================================

Figure \ref{fig6} presents a comparison between the exact Carroll-Hioe solution \eqref{CH} and our approximate solution \eqref{probabilities}.
Quite astonishingly, our approximate solution is not only qualitatively correct but it is even in a very good quantitative agreement with the exact solution;
 we witness here yet another LZ surprise where our LZ-based model outperforms expectations in a limit where it should not be adequate.

The observed feature of our approximate solution can be explained by examining the asymptotics of the approximate probabilities \eqref{probabilities} and the exact CH values \eqref{CH} for $a\ll 1$ and $a\gg 1$.
For $a\ll 1$ the approximation \eqref{probabilities} and the CH solution \eqref{CH} read, up to $\O(a^4)$, respectively
\bse\label{small a}
\bea\label{small a approx}
&&\P \sim \left[ \begin{array}{ccc}
1-5\pi a^2/4 & \pi a^2  & \pi a^2/4  \\
\pi a^2  & 1-2\pi a^2  & \pi a^2  \\
\pi a^2/4 & \pi a^2  & 1-5\pi a^2/4
\end{array}\right],\\
&&\P_{CH} \sim \left[ \begin{array}{ccc}
1-\pi a^2 & \pi a^2  & 0 \\
\pi a^2  & 1-2\pi a^2 & \pi a^2  \\
0 & \pi a^2  & 1-\pi a^2
\end{array}\right].
\label{small a CH}
\eea
\ese
For $a\gg 1$ they read, up to $\O(e^{-\pi a^2})$, respectively
\bse\label{large a}
\bea\label{large a approx}
&&\P \sim \left[ \begin{array}{ccc}
0 & e^{-\pi a^2/4}  & 1-e^{-\pi a^2/4}  \\
0 & 1-e^{-\pi a^2/4}  & e^{-\pi a^2/4}  \\
1 & 0  & 0
\end{array}\right],\\
&&\P_{CH} \sim \left[ \begin{array}{ccc}
0                & 2e^{-\pi a^2/2}   & 1-2e^{-\pi a^2/2} \\
2e^{-\pi a^2/2}   & 1-4e^{-\pi a^2/2} & 2e^{-\pi a^2/2}  \\
1-2e^{-\pi a^2/2} & 2e^{-\pi a^2/2}   & 0
\end{array}\right].
\label{large a CH}
\eea
\ese

Equations \eqref{small a approx} and \eqref{small a CH} demonstrate that our approximate solution \eqref{probabilities} reproduces well, for some probabilities even exactly, the correct small-$a$ asymptotics,
 which corresponds to the large-$A$ ranges in Fig. \ref{fig6}.
The reason is that the small-$a$ (diabatic) regime corresponds to weak coupling; in the perturbative regime the presence of level crossings, let alone their distribution in time, is less significant.
In the large-$a$ (adiabatic) regime the crossings become very important and definitive for the dynamics.
Then Eq. \eqref{large a approx} deviates from the correct asymptotics \eqref{large a CH}, but still has the correct asymptotic values for $a\rightarrow \infty$.
The correct, or nearly correct, small-$a$ and large-$a$ asymptotics of our approximate solution \eqref{probabilities} explain its surprising overall accuracy in Fig. \ref{fig6}.

%%%%%%%%%%%%%%%%%%%%%%%%%%%%%%%%%%%%%%%%%%%%%%%%%%%%%%%%%%%%%%%%%%%%%%%
%%%%%%%%%%%%%%%%%%%%%%%%%%%%%%%%%%%%%%%%%%%%%%%%%%%%%%%%%%%%%%%%%%%%%%%
\section{Discussion and conclusions}\label{Sec-conclusions}
%%%%%%%%%%%%%%%%%%%%%%%%%%%%%%%%%%%%%%%%%%%%%%%%%%%%%%%%%%%%%%%%%%%%%%%
%%%%%%%%%%%%%%%%%%%%%%%%%%%%%%%%%%%%%%%%%%%%%%%%%%%%%%%%%%%%%%%%%%%%%%%

We have developed an approximate analytical model that describes the time-dependent dynamics of a quantum system with three states,
 which have linearly changing energies of different slopes and are coupled with pulse-shaped interactions.
Our approach is based upon the two-state LZ model, i.e. we assume independent pairwise transitions between neighboring states, described by the LZ model.
We have performed detailed comparison of our analytic approximation with numerical simulations, versus all possible interaction parameters and for all nine transition probabilities,
 which has revealed a remarkable accuracy, not only in smooth features, but also in describing detailed interference features.
This accuracy shows that indeed, the physical mechanism of the three-state dynamics is dominated by separated pairwise LZ transitions, even when the crossings are too close to each other.

We have derived the \textit{formal} conditions of validity of our LZ approach, Eqs. \eqref{conditions}, using the concept of transition time.
However, a comparison with numeric simulations has revealed that our approximation is valid well beyond the formal ranges of validity.
One of the reasons is that for two of the survival probabilities, $P_{1\rightarrow 1}$ and $P_{3\rightarrow 3}$, our LZ approximation produces the exact results.
We have found that even in the extreme case of vanishing static detuning, %$\Delta_0\rightarrow 0$,
 where our approach \textit{should not be valid} because the three crossings coalesce into a triply degenerate bowtie single crossing,
 it still produces remarkably accurate results because of nearly correct asymptotic behaviors of the transition probabilities.

One of the useful and interesting features of the ``triangle'' linkage pattern (Fig. \ref{fig1}) is the presence of intrinsic interference effects.
Our ``sandwich'' approach, with its implementation in the adiabatic interaction representation, allows for an easy incorporation of different evolution paths in Hilbert space between a particular pair of states.
Such \textit{path interferences} are identified in only four of the nine probabilities.
Another source of interferences could be nonadiabatic transitions in the wings of the Gaussian pulses, where the nonadiabatic couplings possess local maxima;
 these interferences would be visible in all nine probabilities.
We have found, however, that only the path interferences are clearly identified.

A substantial contribution to the path interferences is played by the LZ phases $\phi_\subscr$.
The LZ phase is often neglected in applications of the LZ model to multiple crossings, in the so-called ``independent crossing'' approximation, where only probabilities are accounted for.
Although such an approach occasionally works, miraculously, as in the exactly soluble Demkov-Osherov \cite{Demkov68} and Demkov-Ostrovsky \cite{Demkov95} models,
 the present simple, but very instructive model, demonstrates that in general, the LZ phase, as well the dynamical adiabatic phases, has to be properly accounted for,
 which is achieved best in an evolution-matrix approach, preferably in the adiabatic-interaction representation \cite{periodic}.

In order to be closer to experimental reality, in the examples we have assumed pulsed interactions, specifically of Gaussian time dependence.
This proved to be no hindrance for the accuracy of the model, which is remarkable because we have applied the LZ model (which presumes constant couplings) at crossings (the first and the last ones)
 situated at the wings of the Gaussian-shaped couplings where the latter change rapidly.
This robustness of the approach can be traced to the use of the adiabatic basis where the pulse-shape details are accounted for in the adiabatic phases.

We have used the analytic results to derive some useful features of the dynamics, for instance, we have found explicitly the parameter values for which certain probabilities reach the 50\% level,
 and for which a maximally coherent superposition is created of all three states $P_1=P_2=P_3=1/3$.

In the specific derivations we have assumed for the sake of simplicity equal couplings for the two transitions and slopes of different signs but equal magnitudes for two of the energies.
These assumptions simplify considerably the ensuing expressions for the probabilities;
 moreover, they are actually present in some important applications (quantum rotors, Zeeman sublevels in magnetic field and spin-spin entanglement).
The formalism is readily extended to the general case, of unequal couplings and different slopes,
 and we have verified that the resulting LZ-based approximation is very accurate again.

To conclude, the present work demonstrates that, once again, the LZ model outperforms expectations when applied to multistate dynamics, with multiple level crossings and a multitude of evolution paths.

%%%%%%%%%%%%%%%%%%%%%%%%%%%%%%%%%%%%%%%%%%%%%%%%%%%%%%%%%%%%%%%%%%%%%%%%
\acknowledgments

This work has been supported by the EU ToK project CAMEL (Grant No. MTKD-CT-2004-014427), the EU RTN project EMALI (Grant No. MRTN-CT-2006-035369),
 and Bulgarian National Science Fund Grants No. WU-205/06 and No. WU-2517/07.

%%%%%%%%%%%%%%%%%%%%%%%%%%%%%%%%%%%%%%%%%%%%%%%%%%%%%%%%%%%%%%%%%%%%%%%
%%%%%%%%%%%%%%%%%%%%%%%%%%%%%%%%%%%%%%%%%%%%%%%%%%%%%%%%%%%%%%%%%%%%%%%
%%%%%%%%%%%%%%%%%%%%%%%%%%%%%%%%%%%%%%%%%%%%%%%%%%%%%%%%%%%%%%%%%%%%%%%
%%%%%%%%%%%%%%%%%%%%%%%%%%%%%%%%%%%%%%%%%%%%%%%%%%%%%%%%%%%%%%%%%%%%%%%
%%%%%%%%%%%%%%%%%%%%%%%%%%%%%%%%%%%%%%%%%%%%%%%%%%%%%%%%%%%%%%%%%%%%%%%

\end{document}